\documentclass[11pt]{article}

\usepackage{amsthm,color,url,epsfig}

\usepackage{amssymb, amsmath, epsfig}
\usepackage{tabularx}
\usepackage{multirow}
\usepackage{tikz}
\usetikzlibrary{shapes}

\newcommand\oo{{\mathcal O}}

\newtheorem{theorem}{Theorem}[section]

\newtheorem{definition}[theorem]{Definition}
\newtheorem{proposition}[theorem]{Proposition}

\begin{document}

\title{Solving Polynomial Systems in the Cloud \\
       with Polynomial Homotopy Continuation\thanks{This material
is based upon work supported by the National Science Foundation under
Grant No.\ 1440534.}}

\author{Nathan Bliss \and Jeff Sommars \and Jan Verschelde \and Xiangcheng Yu}

\date{Department of Mathematics, Statistics, and Computer Science \\
 University of Illinois at Chicago \\
 851 South Morgan (M/C 249) \\
 Chicago, IL 60607-7045, USA}

\maketitle

\begin{abstract}
Polynomial systems occur in many fields of science and engineering.
Polynomial homotopy continuation methods apply symbolic-numeric
algorithms to solve polynomial systems.
We describe the design and implementation of our web interface
and reflect on the application of polynomial homotopy continuation
methods to solve polynomial systems in the cloud.
Via the graph isomorphism problem we organize and classify
the polynomial systems we solved.  The classification with
the canonical form of a graph identifies newly submitted
systems with systems that have already been solved.

\end{abstract}

\section{Introduction}

The widespread availability and use of high speed internet connections
combined with relatively inexpensive hardware enabled cloud computing.
In cloud computing, users of software no longer download and install
software, but connect via a browser to a web site, and interact with
the software through a web interface.  Computations happen at some
remote server and the data (input as well as output) are stored 
and maintained remotely.  Quoting~\cite{RVGSP10}, ``Large, virtualized
pools of computational resources raise the possibility of a new,
advantageous computing paradigm for scientific research.''

This model of computing offers several advantages to the user;
we briefly mention three benefits.
First, installing software can be complicated and a waste of time,
especially if one wants to perform only one single experiment to
check whether the software will do what is desired --- the first
author of~\cite{SSM15} has an account on our web server.
In cloud computing, the software installation is replaced with
a simple sign up, as common as logging into a web store interface.
One should not have to worry about upgrading installed software to
newer versions.
The second advantage is that for computationally intensive tasks,
the web server can be aided by a farm of compute servers.
Thirdly, the input and output files are managed at the server.
The user should not worry about storage, as the web server could 
be aided by file servers.
A good web interface helps to formulate the input problems
and manage the computed results.

In this paper, we describe a prototype of a first web interface
to the blackbox solver of PHCpack~\cite{Ver99}.
This solver applies homotopy continuation methods to polynomial systems.
Its blackbox solver, available as \mbox{\tt phc -b} at the command line,
seems the most widely used and popular feature of the software.
Our web interface is currently running at
{\tt https://kepler.math.uic.edu}.

In addition to the technical aspects of designing a web interface,
we investigate what it means to run a blackbox solver in the cloud.
Because the computations happen remotely, the blackbox solver not
only hides the complexity of the algorithms, but also the actual
cost of the computations.  In principle, the solver could use just
one single core of a traditional processor, or a distributed cluster 
of computers, accelerated with graphics processing units~\cite{VY15a},
\cite{VY15b}.
One should consider classifying the hardness of an input problem
and allocating the proper resources to solve the problem.
This classification problem could be aided by mining a database
of solved problems.

To solve the classification problem we show that the problem of
deciding whether two sets of support sets are isomorphic can be
reduced to the graph isomorphism problem~\cite{MP12}, \cite{MP14}.
Our classification problem is related to the isomorphism of
polynomials problem~\cite{Pat96} in multivariate 
cryptology~\cite{BFV13}, \cite{FP06}.
Support sets of polynomials span Newton polytopes
and a related problems is the polytope isomorphism problem,
see~\cite{KS03} for its complexity, which is as hard as the
graph isomorphism problem.

\section{Related Work and Alternative Approaches}

% In the world of free and open source mathematical software,
The Sage notebook interface and the newer SageMathCloud can be
alternative solutions to setting up a standalone cloud service.
PHCpack is an optional package in Sage~\cite{Sage},
available through the {\tt phc.py} interface, developed
by Marshall Hampton and Alex Jokela, based on the earlier
efforts of Kathy Piret~\cite{Pir08} and William Stein.
We plan to upgrade the existing {\tt phc.py} in Sage with 
{\tt phcpy}~\cite{Ver14}.

The computational algebraic geometry software Macaulay2~\cite{M2}
distributes {\tt PHCpack.m2}~\cite{GPV13}, which is a package that
interfaces to PHCpack.  Macaulay2 runs in the cloud as well.
Below are the input commands to the version of Macaulay2 that
runs online.  The output is omitted.
\begin{verbatim}
Macaulay2, version 1.6

i1 : loadPackage "PHCpack";
i2 : help(PHCpack);
i3 : help solveSystem;
i4 : R = CC[x,y,z];
i5 : S = {x+y+z-1, x^2+y^2, x+y-z-3};
i6 : solveSystem(S)
\end{verbatim}

% \noindent Our future project is to upgrade {\tt PHCpack.m2},
% as PHCpack further improves.

\section{Design and Implementation}

All software in this project is free and open source,
as an application of the LAMP stack, where 
LAMP stands for Linux, Apache, MySQL, and Python.
Our web server runs Red Hat Linux, Apache~\cite{CB04} as the web server,
MySQL~\cite{Dub06} as the database,
and Python as the scripting language.
Our interest in Python originates in its growing ecosystem
for scientific computing~\cite{PGH11}.
In our current implementation we do not take advantage of
any web framework.  Our web interface consists mainly of 
a collection of Python CGI scripts.

We distinguish three components in the development of our web interface:
the definition of the database, the sign up process, and the collection
of Python scripts that are invoked as the user presses buttons.
In the next three paragraphs we briefly describe these three components.

MySQL is called in Python through the module {\tt MySQLdb}.
The database manages two tables.  
One table stores data about the users,
which includes their encrypted passwords and generated random names
that define the locations of the folders with their data on the server.
In the other table, each row holds the information about a
polynomial system that is solved.  In this original setup,
mathematical data are not stored in the database.
Every user has a folder which is a generated random 40-character string.
With every system there is another generated 40-character string.
The Python scripts do queries to the database to locate the data
that is needed.

When connecting to the server, the {\tt index.html} leads directly
to the Python script that prints the first login screen.
The registration script sends an email to the first time user
and an activation script runs when that first time user clicks 
on the link received in the registration email.

The third component of the web interface consists of the Python
scripts that interact with the main executable program,
the {\tt phc} built with the code in PHCpack.
Small systems are solved directly.
Larger systems are placed in a queue that is served by compute servers.

\section{Solving by Polynomial Homotopy Continuation}

When applying polynomial homotopy continuation methods to solve
polynomial systems, we distinguish two different approaches.
The first is the application of a blackbox solver, and
the second is a scripting interface.

\subsection{Running a Blackbox Solver}

For the blackbox solver, the polynomials are the {\em only} input.
The parameters that control the execution options are set to work
well on a large class of benchmark examples; and/or tuned
automatically during the solving.
While the input is purposely minimal, the output should
contain various diagnostics and checks.  In particular, the user
must be warned in case of ill conditioning and nearby singularities.
The form of the output should enable the user to verify (or falsify)
the computed results.

The current blackbox solver {\tt phc -b} was designed 
in~\cite{Ver99} for {\em square} problems, that is, systems with 
as many equations as unknowns.
Polyhedral homotopies~\cite{HS95} \cite{VVC94} 
are optimal for sparse polynomial systems.  
This means that the mixed volume is a sharp root count for
generic problems ({\tt phc -b} calls MixedVol~\cite{GLW05} for a fast
mixed volume computation).  Every path in a polyhedral homotopy ends at
an isolated root, except for systems that have special initial 
forms~\cite{HV98}.  For a survey, see e.g.~\cite{Li03}.

Special cases are polynomials in one variable, linear systems,
and binomial systems.  A binomial system has exactly two monomials
in every polynomial.  Its isolated solutions are determined by
a Hermite normal form of the exponent vectors.  Its positive
dimensional solution sets are monomial maps~\cite{AV13b}.

A more general blackbox solver should operate without any
assumptions on the dimension of the solution sets.
Inspired by tropical algebraic geometry, and in particular its
fundamental theorem~\cite{JMM08}, we can generalize polyhedral
homotopies for positive dimensional solution sets, as was done
for the cyclic $n$-roots problems in~\cite{AV12}, \cite{AV13}.
This general polyhedral method computes tropisms based on the
initial forms and then develops Puiseux series starting at the
solutions of the initial forms.

\subsection{The Scripting Interface {\tt phcpy}}

The Python package {\tt phcpy} replaces the input and output files
by persistent objects.  Instead of the command line interface of {\tt phc}
with its interactive menus, the user of {\tt phcpy} runs Python scripts, or
calls functions from the {\tt phcpy} modules in an interactive Python shell.
Version 0.1.4 of {\tt phcpy} is described in~\cite{Ver14}.

The current version 0.2.5 exports most of the tools needed to
compute a numerical irreducible decomposition~\cite{SVW03}.
With a numerical irreducible decomposition one gets all solutions,
the isolated solutions as well as the positive dimensional sets.
The latter come in the form of as many generic points as the degree
of each irreducible component, satisfying as many random hyperplanes
as the dimension of the component.

\section{Pattern Matching with a Database}

Solving a system of polynomial equations for the first time with
polynomial homotopy continuation happens in two stages.
In the first stage, we construct and solve a simpler system than
the original problem.  This simpler system serves as a start system
to solve the original system in the second stage.
Polynomial homotopy continuation methods deform systems with
known solutions into systems that need to be solved.
Numerical predictor-corrector methods track solution paths from
one system to another.

In many applications polynomial systems have natural parameters
and often users will present systems with the same input patterns.

\subsection{The Classification Problem}

If we solve polynomial systems by homotopy continuation,
we first solve a similar system, a system with the same monomial structure,
but with generic coefficients.  If we could recognize the structure of
a new polynomial system we have to solve, then we could save on the
total solving time of the new polynomial system, because we skip the
first stage of solving the start system.  Furthermore, we could give
a prediction of the time it will take to solve the new system with the
specific coefficients.

Giving a name such as {\tt cyclic n-roots} as a query to a search engine is
likely to come up with useful results because this problem has been
widely used to benchmark polynomial system solvers. 
But if one has only a particular formulation of a polynomial system, 
then one would like to relate the particular polynomials to the collection
of polynomial systems that have already been solved.
Instead of named polynomial systems, we work with anonymous mathematical
data where even the naming of the variables is not canonical.

For example, the systems

\begin{equation}
   \left\{
      \begin{array}{r} 
         x^2 + x y^2 - 3 = 0 \\
         2 x^2 y + 5 = 0 \\
      \end{array}
   \right.
   \quad \mbox{and} \quad
   \left\{
      \begin{array}{r}
         3 + 2 a b^2 = 0 \\
         b^2 - 5 + 2 a^2 b = 0
      \end{array}
   \right.
\end{equation}
must be considered isomorphic to each other.

The support set of a polynomial is the set of all exponent tuples
of the monomials that appear in the polynomial with nonzero coefficient.
For the systems in the specific example above, the sets of support
sets are
\begin{eqnarray}
  & & \left\{ \left\{ (2,0), (1,2), (0,0) \right\},
     \left\{ (2,1),  (0,0) \right\} \right\}  \\
  & \mbox{and} &
   \left\{ \left\{ (0,0), (1, 2) \right\},
     \left\{ (0, 2), (0,0), (2, 1) \right\} \right\}.
\end{eqnarray}

\begin{definition} {\rm 
We say that two sets of support sets are {\em isomorphic} 
if there exists a permutation of their equations and variables
so that the sets of support sets are identical. }
\end{definition}
The problem is then to determine whether the sets of support sets of
two polynomial systems are isomorphic to each other.
This problem is related to the isomorphism of polynomials 
problem~\cite{Pat96}.  Algorithms in multivariate cryptology apply
Gr\"{o}bner basis algorithms~\cite{FP06}
and graph-theoretic algorithms~\cite{BFV13}.

\subsection{The Graph Isomorphism Problem}

If we encode the sets of support sets of a polynomial system
into a graph, then our problem reduces to the graph isomorphism problem,
for which practical solutions are available~\cite{MP14}
and accessible in software~\cite{MP12}. 
The problem of determining whether two sets of support sets are isomorphic is
surprisingly nontrivial.
We begin with some theoretical considerations before moving on 
to implementation.

\begin{definition} {\rm 
The {\em graph isomorphism problem} asks whether for two undirected 
graphs $F,G$ there is a bijection $\phi$ between their vertices that 
preserves incidence--i.e.\ if $a$ and $b$ are vertices connected by an edge in $F$, 
then $\phi(a)$ and $\phi(b)$ are connected by an edge in $G$. }
\end{definition}

\begin{proposition}\label{thm:GIequiv}
The problem of determining whether two sets of support sets are isomorphic 
is equivalent to the graph isomorphism problem.
\end{proposition}

\noindent {\em Proof.}
We will give a constructive embedding in both directions.

$(\supseteq)$ We start by showing that graph isomorphism can be embedded in 
checking isomorphism of sets of support sets.
Recall that the incidence matrix of a 
graph $G=(V,E)$ is a matrix with $\#V$ rows and $\#E$ columns where 
the $(i,j)^{th}$ entry is 1 if the $i^{th}$ vertex and $j^{th}$ edge 
are incident and 0 otherwise. 
It is straightforward to show that two graphs are 
isomorphic if and only if their incidence matrices can be made equal 
by rearranging their rows and columns.

Now suppose we have a graph with incidence matrix $A=(a_{ij})$. 
Construct a polynomial by considering the rows as variables and the 
columns as monomials. To be precise, set
\begin{equation}
   p_A = \sum_j \prod_i x_i^{a_{ij}}.
\end{equation}
For example, if a graph has incidence matrix
\begin{equation}
   A = 
   \left( 
     \begin{array}{ccc}
        1 & 1 \\
        1 & 0 \\
        0 & 1
     \end{array}
   \right), \quad \mbox{then} \quad
   p_A = x_1 x_2 + x_1 x_3. 
\end{equation}
Switching two columns of the matrix corresponds to reordering the monomials 
in the sum; switching two rows corresponds to a permutation of variables. 
Therefore, to determine if two graphs are isomorphic, one may find their 
incidence matrices, form polynomials from them, and check if the polynomials
(thought of as single-polynomial systems) are the same up to permutation of
variables. 
Finally, it is important to note that this is a polynomial time reduction. 
To create the polynomial, start at the first column and iterate through all 
of the columns. 
When done with the first column, we have created the first monomial. Repeat 
this process for every column, touching every entry in the matrix exactly once. 
Given an incidence matrix of a graph $G=(V,E)$ with $\#V=n,\#E=m$, 
converting to a polynomial requires $\oo(nm)$ operations.

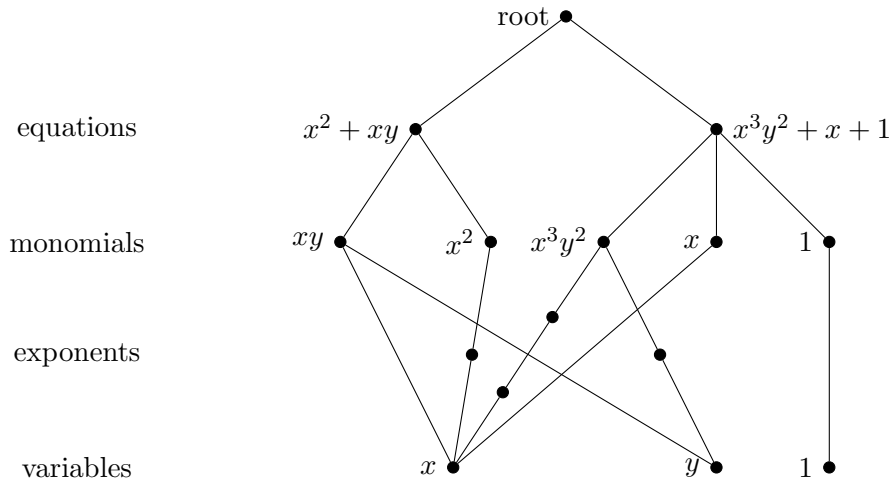
\begin{figure}[hbt]
\centering
\begin{tikzpicture}
\node (root) at (3,6) [label=left:{root},draw, circle,fill,inner sep=1.5pt] {};
\node (f1) at (1,4.5)  [label=left:{$x^2+xy$},draw, circle,fill,inner sep=1.5pt] {};
\node (f2) at (5,4.5)  [label=right:{$x^3y^2+x+1$},draw, circle,fill,inner sep=1.5pt] {};
\draw (root) edge (f1)  (root) edge (f2);

\node (xy) at (0,3) [label=left:{$xy$},draw,circle,fill,inner sep=1.5pt] {};
\node (x2) at (2,3) [label=left:{$x^2$},draw,circle,fill,inner sep=1.5pt] {};
\draw (f1) edge (xy)  (f1) edge (x2);

\node (x3y2) at (3.5,3) [label=left:{$x^3y^2$},draw,circle,fill,inner sep=1.5pt] {};
\node (x) at (5,3) [label=left:{$x$},draw,circle,fill,inner sep=1.5pt] {};
\node (1) at (6.5,3) [label=left:{$1$},draw,circle,fill,inner sep=1.5pt] {};
\draw (f2) edge (x3y2)  (f2) edge (x)  (f2) edge (1);

\node (xvar) at (1.5,0)  [label=left:{$x$},draw,circle,fill,inner sep=1.5pt] {};
\node (yvar) at (5,0) [label=left:{$y$},draw,circle,fill,inner sep=1.5pt] {};
\node (const) at (6.5,0) [label=left:{$1$},draw,circle,fill,inner sep=1.5pt] {};

\draw (xy) edge (xvar);
\draw (xy) edge (yvar);
\node (x2_0) at (1.75,1.5) [draw, circle,fill,inner sep=1.5pt] {};
\draw (x2) edge (x2_0) (x2_0) edge (xvar);
\node (x3_0) at (2.16,1) [draw, circle,fill,inner sep=1.5pt] {};
\node (x3_1) at (2.82,2) [draw, circle,fill,inner sep=1.5pt] {};
\draw (x3y2) edge (x3_1) (x3_1) edge (x3_0) (x3_0) edge (xvar);
\node (y2_0) at (4.25,1.5) [draw, circle,fill,inner sep=1.5pt] {};
\draw (x3y2) edge (y2_0) (y2_0) edge (yvar);
% \node (x_0) at (3.25,1.5) [draw, circle,fill,inner sep=1.5pt] {};
% \draw (x) edge (x_0) (x_0) edge (xvar);
\draw (x) edge (xvar);
\draw (1) edge (const);

\node at (-3.5,4.5) {equations};
\node at (-3.5,3) {monomials};
\node at (-3.5,1.5) {exponents};
\node at (-3.5,0) {variables};
\end{tikzpicture}
\caption{Basic graph from system}
\label{fig:basicGraph}
\end{figure}

$(\subseteq)$ We now show the other direction: checking isomorphism of 
sets of support sets can be embedded in graph isomorphism.  
To do so, we give a way of setting up a graph from a set of support sets,
which is most easily seen via an example.
The graph for the system $\{x^2+xy,x^3y^2+x+1\}$ is shown 
in Figure~\ref{fig:basicGraph}. As can be seen from the diagram, 
we build a graph beginning with a root node to ensure the graph is connected.
We attach one node for each equation, then attach nodes to each equation for
its monomials. Finally we put a row of variable nodes at the bottom and 
connect each monomial to the variables it contains via as many segments 
as the degree of the variable in the monomial.

\begin{figure}[hbt]
\centering
\begin{tikzpicture}
\node (root) at (1,3) [label=left:{root},draw, circle,fill,inner sep=1.5pt] {};
\node (f1) at (0,2)  [label=left:{$x$},draw, circle,fill,inner sep=1.5pt] {};
\node (f2) at (2,2)  [label=right:{$y$},draw, circle,fill,inner sep=1.5pt] {};
\draw (root) edge (f1)  (root) edge (f2);
\node (mon1) at (0,1)  [draw, circle,fill,inner sep=1.5pt] {};
\node (mon2) at (2,1)  [draw, circle,fill,inner sep=1.5pt] {};
\draw (f1) edge (mon1) (f2) edge (mon2);
\node (var1) at (0,-1)  [label=left:{$x$},draw, circle,fill,inner sep=1.5pt] {};
\node (var2) at (2,-1)  [label=right:{$y$},draw, circle,fill,inner sep=1.5pt] {};
\draw (mon1) edge (var1) (mon2) edge (var2);
\node at (-2.5,2) {equations};
\node at (-2.5,1) {monomials};
\node at (-2.5,0) {exponents};
\node at (-2.5,-1) {variables};
\node at (1,-1.7) {$\{x,y\}$};

\node (root) at (4,3) [label=right:{root},draw, circle,fill,inner sep=1.5pt] {};
\node (eq) at (4,2) [label=left:{$x^4$},draw, circle,fill,inner sep=1.5pt] {};
\node (mon) at (4,1)[draw, circle,fill,inner sep=1.5pt] {};
\node (1) at (4,0.5)[draw, circle,fill,inner sep=1.5pt] {};
\node (2) at (4,0)[draw, circle,fill,inner sep=1.5pt] {};
\node (3) at (4,-0.5)[draw, circle,fill,inner sep=1.5pt] {};
\node (var) at (4,-1)[label=left:{$x$},draw, circle,fill,inner sep=1.5pt] {};
\draw (root) edge (var);
\node at (4,-1.7) {$\{x^4\}$};

\end{tikzpicture}
\caption{Isomorphic graphs}
\label{fig:badgraphs}
\end{figure}
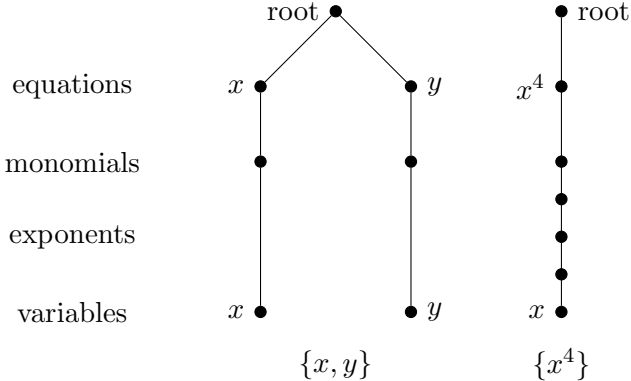

At this point our graph setup is inadequate, as two different systems could
have isomorphic graphs.  Consider the graphs for $\{x,y\}$ and $\{x^4\}$, 
seen in Figure~\ref{fig:badgraphs}. Even though the systems are different, 
the graphs are clearly isomorphic. To remedy this, self-loops are added 
in order to partition the graph into root, equations, monomials, and 
variables by putting one self-loop at the root node, two on each equation 
node, etc. This graph will uniquely represent our system since any
automorphism will be a permutation of nodes within their partitions. 
Partitioning by self-loops is possible because there are no self-loops 
in the initial setup. The graph in Figure~\ref{fig:basicGraph} is drawn 
without the self-loops for the sake of readability.

The time to complete this process is easily seen to be polynomial in the 
number of monomials and the sum of their total degrees. 

Because we have shown that two sets of support sets are isomorphic
if and only if two uniquely dependent graphs 
are isomorphic, and had previously shown that two graphs are isomorphic 
if and only if two uniquely dependent polynomials are equal up to rearranging
variables, we are done. 
Note that this puts our problem in {\bf GI}, 
the set of problems with polynomial time reduction to the graph isomorphism 
problem.~\qed

\subsection{Computing Canonical Graph Labelings With Nauty}

The graph isomorphism problem is one of the few problems which is not 
known to be P or NP-complete~\cite{MP14}. Although there is no known 
worst-case polynomial time algorithm, solutions which are fast in practice 
exist. One such solution is {\tt nauty} \cite{MP12}, 
a program which is able to compute both graph 
isomorphism and canonical labelings. More information on {\tt nauty} 
and other software for solving graph isomorphism can be found in \cite{MP14}.

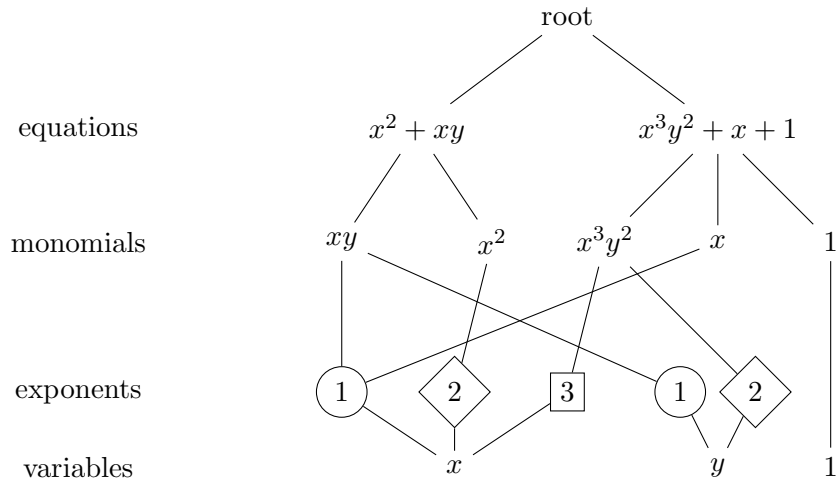
\begin{figure}
\centering
\begin{tikzpicture}
\node (root) at (3,6) [rectangle] {root};
\node (f1) at (1,4.5)  [rectangle]{$x^2+xy$};
\node (f2) at (5,4.5)  [rectangle]{$x^3y^2+x+1$};
\draw (root) edge (f1)  (root) edge (f2);

\node (xy) at (0,3) {$xy$};
\node (x2) at (2,3) {$x^2$};
\draw (f1) edge (xy)  (f1) edge (x2);

\node (x3y2) at (3.5,3) {$x^3y^2$};
\node (x) at (5,3) {$x$};
\node (1) at (6.5,3) {$1$};
\draw (f2) edge (x3y2)  (f2) edge (x)  (f2) edge (1);

\node (xvar) at (1.5,0) {$x$};
\node (yvar) at (5,0) {$y$};
\node (const) at (6.5,0) {$1$};
\node (ex1) [draw,circle] at (0,1) {1};
\node (ex2) [draw,diamond] at (1.5,1) {2};
\node (ex3) [draw,rectangle] at (3,1) {3};
\node (ey1) [draw,circle] at (4.5,1) {1};
\node (ey2) [draw,diamond] at (5.5,1) {2};
\draw (xvar) edge (ex1)  (xvar) edge (ex2)  (xvar) edge (ex3);
\draw (yvar) edge (ey1)  (yvar) edge (ey2);

\draw (xy) edge (ex1)  (xy) edge (ey1);
\draw (x2) edge (ex2);
\draw (x3y2) edge (ex3)  (x3y2) edge (ey2);
\draw (x) edge (ex1);
\draw (1) edge (const);

\node at (-3.5,4.5) {equations};
\node at (-3.5,3) {monomials};
\node at (-3.5,1) {exponents};
\node at (-3.5,0) {variables};
\end{tikzpicture}
\caption{{\tt Nauty} graph setup}
\label{fig:nautygraph}
\end{figure}

Because {\tt nauty} can incorporate the added information of ordered 
partitions, we chose to revise our setup to take advantage of this and 
minimize the number of nodes. Instead of using self-loops to partition 
the graph, we specify equations, monomials, and variables to be three of 
the partitions. We then check for which exponents occur in the system, 
and attach nodes representing these exponents to the variables where 
appropriate. 
Instead of using a sequence of nodes and edges to record the degree as 
we did in the proof, we instead attach monomials to these exponent nodes. 
We group these exponent nodes so that all the nodes representing the lowest 
exponent are in the first partition, all representing the second lowest are
in the second partition, etc.
The setup is shown in Figure~\ref{fig:nautygraph}, where the different 
shapes of the exponent nodes represent the distinct partitions.

If we couple this graph with an ordered list of the exponents that occur in the system, we once again have an object that uniquely represents the isomorphism class of our system. In addition, {\tt nauty} works by computing the automorphisms of a graph and determining one of them to be canonical; the automorphisms must respect the partition, hence partitioning reduces the number of possibilities to check. So not only does this use fewer vertices and edges than our previous setup, but the use of partitions speeds up {\tt nauty}'s calculations.

Another advantage to using {\tt nauty} is that it computes generators of the automorphism group of the graph. Some of these generators permute the equations and monomials. These are unimportant and may be discarded, but the generators that permute the variables are quite useful, as PHCpack has the ability to take advantage of symmetry in polynomial systems. If a system remains the same under certain changes of variables, runtime may be significantly decreased if this symmetry is passed to {\tt phc}.

It is worth noting that by Frucht's theorem \cite{Frucht39} and our 
Proposition~\ref{thm:GIequiv} we immediately obtain that for any 
group $G$ there is a polynomial system with $G$ as its symmetry structure.
If we want to actually find systems with particular structures this is 
fairly impractical, however, as the proof of Frucht's theorem uses the 
Cayley graph which has a node for every group element, meaning that a system built with this method would have a variable for every group element.

We are primarily interested in using {\tt nauty} in the context of storing 
tuples of support sets in a database. Because of this, the canonical labeling
feature is much more useful than the ability to check graph 
isomorphism--looking up a system in a database ought not be done by going 
through a list of systems and querying {\tt nauty} as to whether they are 
isomorphic to the one we are looking up, since this would be highly 
inefficient. Instead we simply parse the system into its graph form 
(including partition data), pass this information to {\tt nauty} 
and compute the canonical labeling, and attach the exponent data, 
as a string, to {\tt nauty}'s output. This process gives us a string 
that uniquely corresponds to the isomorphism class of the system, 
which we can then store in a database.

\subsection{Benchmarking the Canonization}

Timings reported in this section were done 
on a 3.5 GHz Intel Core i5 processor in a iMac Retina 5K
running version 10.10.2 of Mac OS X, with 16 GB RAM.
Scripts are available at
{\tt https://github.com/sommars/PolyGraph}.

In order to design an intelligent storage system, 
it is necessary to know an upper bound of the length of the string. 
As polynomial systems can be arbitrarily large, leading to arbitrarily 
long strings from {\tt nauty}, we chose to analyze a number of well known systems
to act as benchmarks, giving us an idea of an upper bound.
First, consider the cyclic $n$-root polynomial systems. 
These systems consist of $n$ equations: 
$\sum_{i=1}^n \prod_{j=i}^{i+k} x_{j \text{ mod } n} = 0$ for $k$ 
from 0 to $n-2$, and $x_1 \ldots x_n - 1 = 0$. 
For example, the cyclic-3 system is
\begin{equation}
   \left\{
      \begin{array}{r}
          x_1 + x_2 + x_3 = 0~ \\
          x_1x_2 + x_2x_3 + x_3x_1 = 0~ \\
          x_1x_2x_3 - 1 = 0.
      \end{array}
   \right.
\end{equation}

These systems have a lot of symmetry, so they are
an interesting case for us to benchmark this process. 
We tested against both small and large values of $n$ to gain 
a thorough understanding of how this system will be canonized.
Experiments in small dimensions are summarized
in Table~\ref{tabcycsmall}.
The computation time, an average of three trials, is quite fast, 
though the length of the string grows quickly.
In comparison, note that the calculation of the root counts
in the blackbox solver of PHCpack 
(which includes the mixed volume computation)
for the cyclic 10-roots problems takes 48.8 seconds.

\begin{table}[hbt]
\begin{center}
\begin{tabular}{r|r|r|r}
~$n$~ & time~ & ~\#nodes~ & ~\#characters \\ \hline
 4~ & ~0.006~ &  29~~~ &    526~~~ \\ 
 6~ & ~0.006~ &  53~~~ &  1,256~~~ \\ 
 8~ & ~0.006~ &  85~~~ &  2,545~~~ \\ 
10~ & ~0.007~ & 125~~~ &  5,121~~~ \\ 
12~ & ~0.007~ & 173~~~ &  8,761~~~  \\ 
\end{tabular}
\end{center}
\caption{Cyclic $n$-root benchmark for small $n$.
For each $n$, its takes only milliseconds.
We list the number of nodes in the graph and the length
of the canonical form.}
\label{tabcycsmall}
\end{table}

For large instances of cyclic $n$-roots, the exponential growth of the 
number of solutions increases so much that we may no longer hope to
be capable of computing all solutions.
Nevertheless, with GPU acceleration~\cite{VY15}, we can manage to track
a limited number of paths.
Table~\ref{tabcyclarge} illustrates the relationship between the
dimension~$n$, the time, and the sizes of the data. 

\begin{table}[hbt]
\begin{center}
\begin{tabular}{r|r|r|r}
$n$~ &  time~ & ~\#nodes~ & \#characters \\ \hline
 16~ &   0.010~ &    293~~ &     20,029~ \\
 32~ &   0.045~ &  1,093~~ &    168,622~ \\ 
 48~ &   0.265~ &  2,405~~ &    601,702~ \\ 
 64~ &   1.200~ &  4,229~~ &  1,427,890~ \\ 
 80~ &   4.316~ &  6,565~~ &  2,778,546~ \\ 
 96~ & ~15.274~ &  9,413~~ &  4,784,390~ \\ 
112~ & ~38.747~ & 12,773~~ &  8,595,408~ \\ 
128~ & ~80.700~ & 16,645~~ & 13,094,752~ \\   
\end{tabular}
\end{center}
\caption{For larger values of the dimension of the cyclic $n$-root problem,
times and sizes of the data start to grow exponentially.}
\label{tabcyclarge}
\end{table}

Another interesting class of benchmark polynomial systems that we can
formulate for any dimension is the computation 
of Nash equilibria~\cite{Dat09}.
We use the formulation of this problem as in~\cite{MM97}.
The number of isolated solutions also grows exponentially in~$n$.
Table~\ref{tabnash} summarizes our computational experiments.
% corresponds to running genericnashsystem(n) 
% from phcpy for given values of $n$.
As before, the times represent an average of three trials.

\begin{table}[htb]
\begin{center}
\begin{tabular}{r|r|r|r}
$n$~ &    time~~ & \#nodes~ & \#characters \\ \hline
  4~ &    0.006~ &     47~ &     977~  \\
  5~ &    0.006~ &     98~ &     2,325~  \\ 
  6~ &    0.007~ &    213~ &     7,084~  \\ 
  7~ &    0.013~ &    472~ &    18,398~  \\ 
  8~ &    0.054~ &  1,051~ &    51,180~  \\ 
  9~ &    0.460~ &  2,334~ &   134,568~  \\ 
 10~ &    4.832~ &  5,153~ &   331,456~  \\ 
 11~ &   73.587~ & 11,300~ &   872,893~  \\   
 12~ & ~740.846~ & 24,615~ & 2,150,512~  \\  
\end{tabular}
\end{center}
\caption{The cost of the canonization for the Nash equilibria 
polynomial systems for increasing dimension~$n$, 
with the running time expressed in seconds,
the number of nodes, and the size of the canonical form.  }
\label{tabnash}
\end{table}

Compared to the cyclic $n$-roots problem, the dimensions
in Table~\ref{tabnash} are just as small as in Table~\ref{tabcycsmall},
but the time and sizes grow much faster for the Nash equilibria than
for the cyclic $n$-roots.  We suspect two factors.
First, while structured, the Nash equilibria systems are not as sparse
as the cyclic $n$-roots problems.  Second, unlike the cyclic $n$-roots
problem every equation in the Nash equilibria system has the same
structure, so the full permutation group leaves the sets of support
sets invariant.

We end with a system formulated by S.~Katsura,
see~\cite{BGK86} and~\cite{Kat90}.
Table~\ref{tabkatsura} shows the cost of the canonization
of this system.
Because there is no symmetry in the support sets,
the cost of the canonization increases not as fast in
the dimension~$n$ as with the other two examples.

\begin{table}[h!]
\begin{center}
\begin{tabular}{r|r|r|r}
  $n$~ &    time~~ & \#nodes~ & \#characters \\ \hline
   25~ &  ~0.020~ &     929~ &     24,906~  \\
   50~ &  ~0.090~ &     3,411~ &     112,654~  \\ 
   75~ &  ~0.546~ &    7,454~ &     254,770~  \\ 
  100~ &  ~1.806~ &   13,061~ &    495,612~  \\ 
  125~ &  ~4.641~ &  20,229~ &    793,662~  \\ 
  150~ &  ~10.860~ &  28,961~ &   1,157,498~  \\ 
  175~ &  ~21.194~ &  39,254~ &   1,587,115~  \\ 
  200~ &  ~52.814~ & 51,111~ &   2,082,562~  \\   
  225~ &  ~98.118~ & 64,529~ &   2,643,891~  \\  
\end{tabular}
\end{center}
\caption{The cost of the canonization of the Katsura system
for increasing dimension~$n$.
The number of solutions for the $n$-dimensional version of
the system equals~$2^{n}$.}
\label{tabkatsura}
\end{table}

The actual cost of the computation of the canonical form
may serve as an initial estimate on the cost of solving the system.

\subsection{Storing Labelings in a Database}

There are many different ways to design a database to store a given set 
of information.
We do not contend that this is necessarily the best way, 
but it is certainly an effective way of storing our uniquely 
identifying information that will lead to fast lookup. Consider
the explicitly described schema in Figure~\ref{fig:polydatabase}.

\begin{figure}[hbt]
% \subsubsection{Basic database structure by polynomial graph}
\begin{center}
     \begin{tabularx}{\textwidth}{ c | l | c | X}
     \hline
     Section &  Name & Datatype & Description \\ \hline
     \multirow{4}{*}{Graph Nodes}  & $n\_node\_variable$   & INT          & Number of variables nodes  \\ 
                                                    & $n\_node\_monomial$ & INT         & Number of monomial nodes  \\ 
                                                    & $n\_node\_equation$   & INT         & Number of equation nodes  \\ 
                                                    & $n\_node\_degree$     & INT          & Number of degree nodes  \\ \hline
     \multirow{2}{*}{Degree Set}     & $n\_degree$               & INT          & Sum of all degrees  \\
                                                    & $degrees$                   & VARCHAR & Set of all degrees  \\ \hline
     \multirow{3}{*}{Graph}           & $graph\_length$      & INT          & Length of complete graph \\
                                                        & $graph\_filename$   & VARCHAR  &  Complete graph in file \\ \hline
     \multirow{1}{*}{Polynomial Info}   & $poly\_filename$       & VARCHAR   & Information and Reference file for polynomial system  \\ \hline
     \end{tabularx}
\end{center}
\caption{Database structure for polynomial system graph}
\label{fig:polydatabase}
\end{figure}

Each of the data elements in it are used to partition the database 
from the previous elements,
so it can be seen as a B-tree structure. 
For identifying whether or not a set of support sets
is already in our database, 
this would lead to a search time of $O(\log n)$~\cite{Com79}.

\section{Conclusions}

The practical considerations of offering a cloud service to solve
polynomial systems with polynomial homotopy continuation led to
the classification problem of polynomial systems.
To solve this problem, we linked the isomorphism problem for
sets of support sets to the graph isomorphism problem and applied
the software {\tt nauty}.

Although no polynomial time algorithm are known to solve 
the graph isomorphism problem, we presented empirical results from
benchmark polynomial systems that the computation of a canonical form
costs much less than solving a polynomial system.

% We presented a way of using databases to save future computation time, 
% but we intend to work on another way of saving computations. For some
% polynomial systems, computing solutions is sped up drastically through
% the use of tropical methods; however, the first step tends to be finding 
% the tropical prevariety for the system. 
% Finding a tropical prevariety first requires finding the pretropisms of the 
% Newton polytopes of the supports of all of the polynomials. 
% We intend to store pretropisms in a database, 
% saving us from future time consuming repeated computation,
% and ultimately speeding up the computation of solutions of polynomial systems.

% \newpage

\section{Acknowledgement}
The computer that hosts our cloud service was purchased through
a UIC LAS Science award.

\bibliographystyle{plain}
% \bibliography{phcweb}

\end{document}